\documentclass[runningheads]{chaos2009}

\usepackage{chaos2009References} 

\usepackage{graphicx} 

\begin{document}

\title*{QUANTUM SCATTERING AND TRANSPORT IN CLASSICALLY CHAOTIC CAVITIES: \\
An Overview of Past and New Results}
%
\toctitle{QUANTUM SCATTERING AND TRANSPORT IN CLASSICALLY CHAOTIC CAVITIES:
An Overview of Past and New Results}
%
\titlerunning{QUANTUM CHAOTIC SCATTERING}
%
\author{
  Pier A. Mello\inst{1}
  \and 
  V\'ictor A. Gopar\inst{2}
\and
J. A. M\'endez-Berm\'udez\inst{3}
}
%
\index{P. A. Mello}
\index{V. A. Gopar}
\index{J. A. M\'endez-Berm\'udez}

%
\authorrunning{Mello, Gopar and M\'endez}
%
\institute{
Instituto de F\'{\i}sica \\ 
Universidad Nacional Aut\'{o}noma de M\'{e}xico \\ 
01000 M\'{e}xico Distrito Federal, Mexico \\
 (e-mail: {\tt mello@fisica.unam.mx})
\and
Departamento de F\'isica Te\'orica \\ 
and Instituto de Biocomputaci\'on y F\'isica de Sistemas Complejos \\ 
Universidad de Zaragoza \\ 
Pedro Cerbuna 12, 50009 Zaragoza, Spain\\
(e-mail: {\tt gopar@unizar.es})
\and
Instituto de F\'{\i}sica \\ 
Universidad Aut\'onoma de Puebla \\
Apdo. Postal J-48,  Puebla 72570, Mexico \\
(e-mail: {\tt antonio.ifuap@gmail.com})
}

\maketitle             



\begin{abstract}
We develop a statistical theory that describes quantum-mechanical scattering
of a particle by a cavity when the geometry is such that the classical
dynamics is chaotic.
This picture is relevant to a variety of physical systems, ranging from atomic nuclei to mesoscopic systems and microwave cavities;
the main application to be discussed in this contribution is to electronic transport through mesoscopic ballistic structures or quantum dots.
The theory describes the regime in which there are two distinct time scales, associated with a prompt and an equilibrated response, and is cast in terms of the matrix of scattering amplitudes $S$.
We construct the ensemble of $S$ matrices using a maximum-entropy approach which incorporates the requirements of flux conservation, causality and ergodicity, and the system-specific average of $S$ which quantifies the effect of prompt processes.
The resulting ensemble, known as Poisson's kernel, is meant to describe those situations in which any other information is irrelevant.
The results of this formulation have been compared with the numerical solution of the Schr\"{o}dinger equation for cavities in which the
assumptions of the theory hold.
The model has a remarkable predictive power:
it describes statistical properties of the quantum conductance of quantum dots,
like its average, its fluctuations, and its full distribution in several cases.
We also discuss situations that have been found recently, in which the notion of stationarity and ergodicity is not fulfilled, and yet Poisson's kernel gives a good description of the data. At the present moment we are unable to give an explanation of this fact.
\keyword{Quantum Chaotic Scattering}
\keyword{Statistical $S$ matrix}
\keyword{Mesoscopic Physics}
\keyword{Information Theory}
\keyword{Maximum Entropy}
\end{abstract}

\section{Introduction}
\label{intro}

The problem of coherent multiple scattering of waves has long been of great interest in physics and, in particular, in optics.
There are a great many wave-scattering problems, appearing in various fields of physics, where the interference pattern due to multiple scattering is so complex, that a change in some external parameter changes it completely and, as a consequence, only a statistical treatment is feasible and, perhaps, meaningful \cite{mello_kumar_oup,sheng}.

In the problems to be discussed here, complexity in wave scattering derives from the chaotic nature of the
underlying classical dynamics. Our discussion will find applications to problems like:
i) electronic transport in microstructures called ballistic quantum dots, and
ii) transport of electromagnetic waves, or other classical waves (like elastic waves), through cavities with a chaotic classical dynamics.
In particular, we shall study the statistical fluctuations of transmission and reflection of waves, which are of considerable interest in mesoscopic physics.

The ideas involved in our discussion have a great generality.
Let us recall that, historically, nuclear physics
--a complicated many-body problem-- has offered very good examples of complex quantum-mechanical scattering. Here, the typical dimensions are on the scale of the fm ($1$ fm = $10^{-13}$cm). The statistical theory of nuclear reactions has been of great interest for many years, in those cases where, due to the presence of many resonances, the cross section is so complicated as a function of energy that
its detailed structure is of little interest and a statistical treatments is then called for.
Most remarkably, one finds similar statistical properties in the quantum-mechanical scattering of
``simple" one-particle systems --like a particle scattering from a cavity-- whose classical dynamics is chaotic. The typical dimensions of the ballistic quantum dots that were mentioned above is 1 $\mu $m, while those of microwave cavities is of the order of 0.5 meters.
Thus the size of these systems spans $\approx 14$ orders of magnitude!

The purpose of this contribution is to review past and recent work in which various
ideas that were originally developed in the framework of nuclear physics, like
the nuclear optical model and the statistical theory of nuclear reactions
\cite{feshbach},
have been used to give a unified treatment of quantum-mechanical scattering in simple one-particle systems in which the corresponding classical dynamics is chaotic, and of microwave propagation through similar cavities (see Ref. \cite{mello_kumar_oup} and references contained therein).

The most remarkable feature that we shall encounter is the statistical regularity of the results, in the sense that they will be expressible in terms of a few relevant physical parameters, the remaining details being just ``scaffoldings".
This feature will be captured within a framework that we shall call the ``maximum-entropy approach".

In the next section we present some general physical ideas related to scattering of microwaves by cavities
and of electrons by ballistic quantum dots.
The formulation of the scattering problem is given in Sec. \ref{scattering}, where we employ, to be specific, the language of electron scattering.
The notion of doing statistics with the scattering matrix and the resulting maximum-entropy model are reviewed in Secs. \ref{statistical-S}  and \ref{max-entropy}, respectively.
In Sec. \ref{PK-and-numerics}
we compare the predictions of our model with a number of computer simulations. We also present a critical discussion of the results, based on some recent findings on the the validity conditions of our statistical model.
The conclusions of this presentation are the subject of
Sec. \ref{conclusions}.

\section{Microwave cavities and ballistic quantum dots}
\label{cavities}

\begin{figure}
  \centerline{
    \includegraphics[width=.8\textwidth]
    {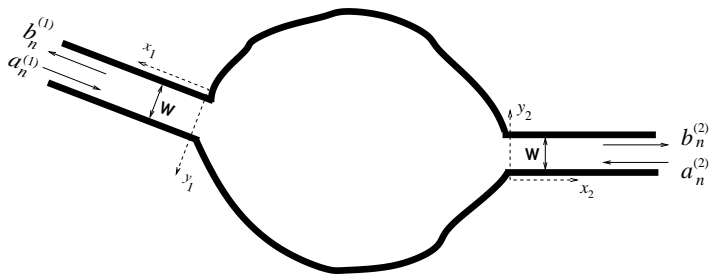}}
  \caption{The 2D cavity referred to in the text.
The cavity is connected to the outside via two waveguides, each supporting $N$
running modes.
The arrows inside the waveguides indicate incoming or outgoing waves.
In waveguide $l=1,2$ the wave amplitudes are indicated as
$a_n^{(l)}$, $b_n^{(l)}$,
respectively, where $n=1,\cdot \cdot \cdot ,N$.
}
  \label{cavity}
\end{figure}

Consider the system shown schematically in Fig. \ref{cavity}.
It consists of a cavity connected to the external world by two waveguides, ideally of infinite length.

We may think of performing a {\em microwave} experiment with such a device \cite{microwave}.
A wave is sent in from one of the waveguides, the result being a reflected wave in the same waveguide and a transmitted wave in the other.
One given point inside the system receives the contributions from many {\em multiply scattered waves} that have bounced from the inner surface of the cavity, giving rise to the complex interference pattern mentioned in the Introduction.
An alternative interpretation of the same situation is that the outgoing wave has suffered the effect of many resonances that are present inside the cavity.
The net result is an interference pattern which shows an appreciable sensitivity to changes in external parameters: for example, a small variation in the shape of the cavity may change the pattern drastically.

In experiments on {\em electronic transport} \cite{expts}
performed with ballistic microstructures, or quantum dots, one finds, for sufficiently low temperatures and for
spatial dimensions on the order of 1$\mu $m or less, that the phase coherence
length $l_{\phi }$ exceeds the system dimensions; thus the phase of the
single-electron wave function --in an independent-electron approximation--
remains coherent across the system of interest. Under these conditions,
these systems are called {\it mesoscopic}
\cite{altshuler,carlo(review),carlo(rmp)}.
The elastic mean free path $l_{el} $ also exceeds the system dimensions:
impurity scattering can thus be neglected, so that only scattering from the boundaries of the system is important.
In these systems the dot acts as a resonant cavity and the leads as electron waveguides.
Experimentally, an electric current is established through the leads that
connect the cavity to the outside, the potential difference is measured
and the conductance $G$ is then extracted.
Assume that the microstructure is placed between two reservoirs (at
different chemical potentials) shaped as expanding horns with negligible
reflection back to the microstructure.
In a picture that takes into account the electron-electron interaction in the
random-phase approximation ({\it RPA}), the two-terminal conductance at zero
temperature --understood as the ratio of the current to the potential difference
between the two reservoirs-- is then given by Landauer's formula \cite{landauer,markus,mello_kumar_oup}
\begin{equation}
G =2\frac{e^{2}}{h}T ,
\label{landauer a}
\end{equation}
where the total transmission $T$ is given by
\begin{equation}
T =tr\left( tt^{\dagger }\right) ,
\label{T}
\end{equation}
$t$ being the matrix of transmission amplitudes $t_{ab}$
associated with the resulting self-consistent single-electron problem at the Fermi energy
$\epsilon _F$, with $a$ and $b$ denoting final and initial channels.
The factor of 2 in Eq. (\ref{landauer a}) is due to the two-fold spin degeneracy in the absence of spin-orbit scattering that we shall assume here.
In this ``scattering approach'' to electronic transport, pioneered by Landauer \cite{landauer}, the problem is thus reduced to understanding the quantum-mechanical single-electron scattering by the cavity under study.

When the system is immersed in an external magnetic field $B$ and the latter is varied, or the Fermi energy $\epsilon _F$ or the shape of the
cavity are varied, the relative phase of the various partial waves changes
and so do the interference pattern and the conductance. This sensitivity of $%
G$ to small changes in parameters through quantum interference is called
{\it conductance fluctuations}.

The experiments on quantum dots given in Ref. \cite{expts}
report cavities in the shape of a stadium, for which the
single-electron classical dynamics would be chaotic, as well as experimental
{\it ensembles} of shapes. The average of the conductance, its fluctuations and
its full distribution were obtained over such ensembles.

In what follows we shall phrase the problem in the language of electronic scattering; spin will be ignored, so that we shall deal with a scalar wave equation.
However, our formalism is applicable to scattering of microwaves, or
other classical waves, in situations where the scalar wave equation is a reasonable approximation.

\section{The scattering problem}
\label{scattering}

In our application to mesoscopic physics we consider a system of
noninteracting ``spinless'' electrons and study the scattering of an
electron at the Fermi energy $\epsilon _{F}$ by a 2D microstructure,
connected to the outside by two leads of width $W$
(see Fig. \ref{cavity}).

Very generally, a quantum-mechanical scattering problem is described by its scattering matrix, or $S$ matrix, that we now define.
The confinement of the electron in the transverse direction in the leads gives rise to the so called  {\em transverse modes}, or channels.
If the incident Fermi momentum $k_{F}$ satisfies
\begin{equation}
N < k_{F}W/\pi < N+1 ,
\label{N}
\end{equation}
there are $N$ transmitting or running modes (open channels) in each of the two leads.
The wavefunction in lead $l$
($l=1,2$ for the left and right leads, respectively)
is written as the $N$-dimensional vector
\begin{equation}
\Psi ^{(l)}(x_l)=[\psi _{1}^{(l)}(x_l),\cdot \cdot \cdot ,\psi _{N}^{(l)}(x_l)]^{T},
\label{spinor}
\end{equation}
the $n$-th component ($n=1, \dots, N$) being a linear combination of plane waves with constant flux (given by $h$), i.e.,
\begin{equation}
\psi _{n}^{(l)}(x_l)
=a_{n}\frac{{\rm e}^{-ik_{n}x_l}}{(2\pi \hbar^2 k_{n}/m)^{1/2}}
+b_{n}\frac{{\rm e}^{ik_{n}x_l}}{(2\pi \hbar^2 k_{n}/m)^{1/2}}\ ,\;\; n=1,\cdot \cdot \cdot ,N,
\label{psichanneln}
\end{equation}
$m$ being the electron mass.
In (\ref{psichanneln}), $k_{n}$ is the ``longitudinal'' momentum in channel $n$, given by
\begin{equation}
k_{n}^{2}+\left[ \frac{n\pi }{W}\right] ^{2}=k_{F}^{2}.
\label{transmom}
\end{equation}
The $2N$-dimensional $S$ matrix relates the incoming to the outgoing
amplitudes as
\begin{equation}
\left[
\begin{array}{c}
b^{(1)} \\
b^{(2)}
\end{array}
\right] =S\left[
\begin{array}{c}
a^{(1)} \\
a^{(2)}
\end{array}
\right] ,
\label{S}
\end{equation}
where $a^{(1)}$, $a^{(2)}$, $b^{(1)}$, $b^{(2)}$ are $N$-dimensional vectors.
The matrix $S$ has the structure
\begin{equation}
S=\left(
\begin{array}{cc}
r & t^{\prime } \\
t & r^{\prime }
\end{array}
\right) ,
\label{S(rt)}
\end{equation}
where $r,t$ are the $N\times N$ reflection and transmission matrices for particles from the left and $r^{\prime },t^{\prime }$ for those from the right:
this thus defines the transmission matrix $t$ used earlier in
Eq. (\ref{T}).

The requirement of {\it flux conservation (FC)} implies
{\em unitarity} of the $S$ matrix \cite{mello_kumar_oup}, i.e.,
\begin{equation}
SS^{\dagger }=I.
\label{unitarity}
\end{equation}
This is the only requirement in the absence of other symmetries.
This is the so-called {\em unitary} case, also denoted by $\beta=2$ in the literature on random-matrix theory.
For a {\it time-reversal-invariant} (TRI) problem (as is the case in the absence of a magnetic field) and no spin, the $S$ matrix, besides being unitary, is
{\em symmetric} \cite{mello_kumar_oup}:
\begin{equation}
S=S^{{T}}.
\label{symmetry}
\end{equation}
This is the {\em orthogonal} case, also denoted by $\beta =1$.
The {\em symplectic case} ($\beta =4$) arises in the presence of half-integral spin and time-reversal invariance, and in the absence of other symmetries, but will not be considered here.

\section{Doing statistics with the scattering matrix}
\label{statistical-S}

\begin{figure}
  \centerline{
    \includegraphics[width=.7\textwidth]
    {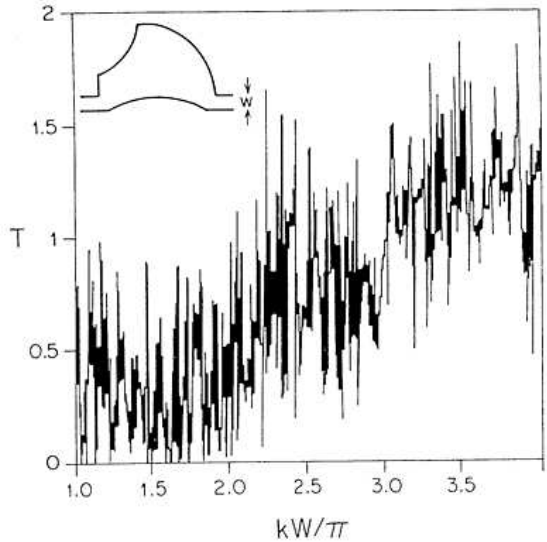}}
  \caption{Total transmission coefficient of Eq. (\ref{T}) as a function of the incident momentum, obtained by solving numerically the Schr\"odinger equation for the system indicated in the upper left corner (from Ref. \protect\cite{pier_harold_physA}), whose classical dynamics has a chaotic nature.}
  \label{numte}
\end{figure}

In numerical simulations of quantum scattering by $2D$ cavities
with a chaotic classical dynamics one finds that the $S$ matrix and hence the various transmission and reflection coefficients fluctuate considerably as a function of the incident energy $E$ (or incident momentum $k$), because of the resonances occurring inside the cavity \cite{mello_kumar_oup}.
These resonances are moderately overlapping for just one open channel and become more overlapping as more channels open up.
An example is illustrated in Fig. \ref{numte}, 
which shows the total transmission coefficient $T(E)$ of Eq. (\ref{T}) [proportional to the conductance, by Landauer's formula of Eq. (\ref{landauer a})],
obtained by solving numerically the Schr\"odinger equation for a cavity whose underlying classical dynamics is chaotic.
Since we are not interested in the detailed structure of $T(E)$, it is clear that we are led to a {\em statistical description} of the problem, just as was explained in the Introduction.

As $E$ changes, $S(E)$ wanders on the unitarity sphere, restricted by the condition of symmetry if TRI applies.
The question we shall address is: what fraction of the ``time" (energy) do we find $S$ inside the ``volume element" $d\mu (S)$?
[$d\mu (S)$ is the invariant measure for $S$ matrices, to be defined below.]
We call $dP(S)$ this fraction and write it as
\begin{equation}
dP(S)= p(S) d\mu (S),
\label{dP_vs_p}
\end{equation}
where $p(S)$ will be called the ``probability density".
Once we phrase the problem in this language, we realize that we are dealing with a {\em random S-matrix theory}.
In the next section we shall introduce a model for the probability density $p(S)$.

The concept of invariant measure $d\mu (S)$ for $S$ matrices mentioned above is an important one and we now briefly describe it.
By definition, the invariant measure for a given symmetry class does not change under an automorphism of that class of matrices onto itself
\cite{dyson,hua}, i.e.,
\begin{equation}
{\rm d}\mu ^{(\beta )}(S)={\rm d}\mu ^{(\beta )}(S^{\prime }).
\label{invmeasure}
\end{equation}
For unitary and symmetric $S$ matrices we have
\begin{equation}
S^{\prime }=U_{0}SU_{0}^{T},
\label{automorphismbeta1}
\end{equation}
$U_{0}$ being an arbitrary, but fixed, unitary matrix.
Eq. (\ref{automorphismbeta1}) represents an automorphism of the set of unitary symmetric matrices onto itself
and defines the invariant measure for the {\em orthogonal} ($\beta=1$) symmetry class of $S$ matrices. We recall that this class is relevant for systems when we ignore spin and in the presence of TRI.
When TRI is broken, we deal with unitary, not necessarily symmetric, $S$ matrices.
The automorphism is then
\begin{equation}
S^{\prime }=U_{0}SV_{0},
\label{automorphismbeta2}
\end{equation}
$U_{0}$ and $V_{0}$ being now arbitrary fixed unitary matrices.
For this {\it unitary}, or $\beta =2$, symmetry class of $S$ matrices the resulting measure is the well known Haar measure of the unitary group and its uniqueness is well known.
Uniqueness for the $\beta =1$ class
(and $\beta =4$, not discussed here) was shown in Ref. \cite{dyson}.
The invariant measure, Eq. (\ref{invmeasure}), defines the so-called {\it Circular (Orthogonal, Unitary) Ensemble (COE, CUE)} of $S$ matrices, for $\beta =1,2$, respectively.

\section{The maximum-entropy model}
\label{max-entropy}

We shall assume that in the scattering process by our cavities
{\em two distinct time scales}
occur \cite{mello_kumar_oup,feshbach,mello-pereyra-seligman,harold-pier3,wrm,lh}:
i) a {\it prompt} response arising from short paths;
ii) a {\it delayed} response arising from very long paths.
Historically, this assumption was first made in the nuclear theory of the compound nucleus, where the prompt response is associated with the so called direct processes, and the delayed one with an equilibrated response arising from the compound-nucleus formation.

The prompt response is described mathematically in terms of $\bar{S}$,
the {\it average}  of the actual scattering matrix $S(E)$ over an energy interval around a given energy $E$. Following the jargon of nuclear physics, we shall refer to $\bar{S}$ as the {\em optical} $S$ matrix.
The resulting averaged, or optical, amplitudes vary much more slowly with energy than the original ones.

As explained below, the statistical distribution of the scattering matrix $S(E)$ in our chaotic-scattering problem is constructed through a maximum-entropy ``ansatz", assuming that it depends parametrically solely on the optical matrix
$\bar{S}$, the rest of details being a mere ``scaffolding".

As in the field of statistical mechanics, it is convenient to think of an ensemble of macroscopically identical cavities, described by an
{\em ensemble of $S$ matrices} (see also Refs. \cite{harold-pier1,carlo_et_al}).
In statistical mechanics, time averages are very difficult to
construct and hence are replaced by {\it ensemble averages} using the notion
of {\em ergodicity}.
In a similar vein, in the present context we idealize $S(E)$, for all real $E$, as a {\em stationary random-matrix function} of $E$ satisfying the condition of
{\em ergodicity} \cite{mello_kumar_oup,lh},
so that we may study energy averages in terms of ensemble averages.
For instance, the optical matrix $\bar{S}$ will be calculated as an ensemble average, i.e., $\bar{S} = \langle S \rangle$, which will also be referred to as the optical $S$ matrix.

We shall assume $E$ to be far from thresholds: locally, $S(E)$ is then a meromorphic function which is analytic in the upper half of the complex-energy plane and has resonance poles in the lower half plane.

From {\em ``analyticity-ergodicity"} one can show the {\em ``reproducing"} property:
given an ``analytic function" of $S$, i.e.,
\begin{equation}
f(S) = \sum_{a_1 b_1 \cdots a_p b_p} c_{a_1 b_1 \cdots a_p b_p}
\;S_{a_1 b_1} \cdots S_{a_p b_p},
\label{analytic f(S)}
\end{equation}
we must have:
\begin{equation}
\langle f(S) \rangle
\equiv \int f(S) \; p_{\langle S \rangle}(S) \; d\mu (S)
=f(\langle S \rangle) .
\label{repr_property}
\end{equation}
This is the mathematical expression of the physical notion of
{\em analyticity-ergodicity}.
Thus $p_{\langle S \rangle}(S)$ must be a {\em ``reproducing kernel"}.
Notice that only the optical matrix $\langle S \rangle$ enters the definition.
The reproducing property (\ref{repr_property}) and reality of the distribution do not fix the ensemble uniquely. However, among the real reproducing ensembles,
{\em Poisson's kernel} \cite{hua}, i.e.,
\begin{equation}
p_{\langle S\rangle }^{(\beta )}(S)
=V_\beta ^{-1}
\frac
{[{\rm det}(I-\langle S\rangle \langle S^{\dagger }\rangle )]
^{(2N \beta + 2-\beta )/2}}
{\mid {\rm det}(I-S\langle S^{\dagger }\rangle )\mid ^{2N \beta + 2-\beta }},
\label{poisson}
\end{equation}
(recall that $2N$ is the dimensionality of the $S$ matrices, $N$ being the number of open channels in each lead)
is special, because its information entropy
\cite{mello_kumar_oup,mello-pereyra-seligman}
\begin{equation}
{\cal S}[p_{\langle S\rangle }]\equiv -\int p_{\langle S\rangle }(S)\;
\ln[p_{\langle S\rangle }(S)]\;d\mu (S)
\label{information}
\end{equation}
is greater than or equal to that of any other probability density satisfying the reproducibility requirement for the same optical $\langle S\rangle $.
We could describe this result qualitatively saying that, for Poisson's kernel, $S$ is {\em as random as possible, consistent with $\langle S\rangle $ and the reproducing property}.

As for its information-theoretic content, Poisson's kernel of Eq. (\ref{poisson}) describes a system with:
I.- the {\it general properties} associated with
i) unitarity of the $S$ matrix (flux conservation),
ii) analyticity of $S(E)$ implied by causality,
iii) presence or absence of time-reversal invariance (and spin-rotation
symmetry when spin is taken into account) which determines the
universality class
[orthogonal ($\beta=1$), unitary ($\beta=2$) or
symplectic ($\beta=4$)],
and II.- the {\em system-specific properties} --parametrized by
the ensemble average
$\left\langle S\right\rangle$--,
which describe the presence of short-time processes.
System-specific {\em details other than the optical
$\left\langle S\right\rangle$ are assumed to be
irrelevant}.

\section{Comparison of the information-theoretic model with numerical simulations}
\label{PK-and-numerics}

A number of computer simulations have been performed, in which the Schr\"{o}dinger equation for 2D structures was solved numerically and the results were compared with our theoretical predictions.
In Refs. \cite{mello_kumar_oup,harold-pier3,wrm,harold-pier1}),
statistical properties of the quantum conductance of
cavities were studied, like its average, its fluctuations and its full
distribution, both in the absence and in the presence of a prompt response.
Here we concentrate on the conductance distribution, mainly in the presence of prompt processes.

\begin{figure}
  \centerline{
    \includegraphics[width=1.0\textwidth]
    {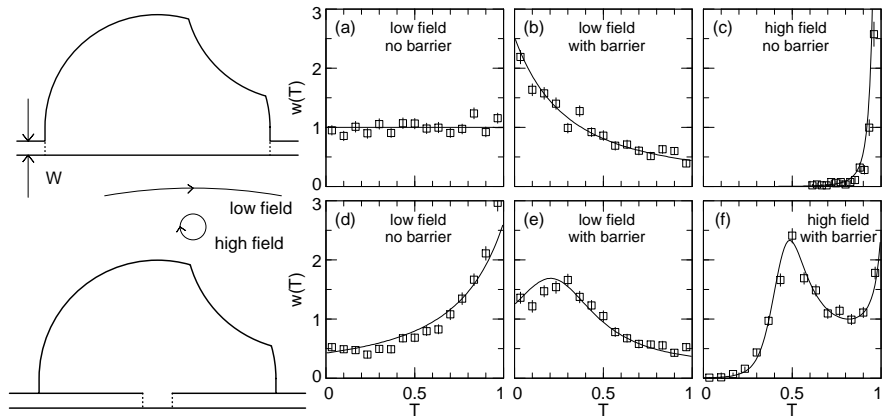}}
  \caption{The probability distribution of the total transmission coefficient $T$, or conductance, obtained by numerically integrating the Schr\"odinger equation for the billiards shown on the left side of the figure for $N=1$. The results are indicated with squares, that include the statistical error bars.
The curves are obtained from Poisson's kernel, Eq. (\ref{poisson}), with
$\left\langle S \right\rangle$ extracted from the numerical data. The agreement is good.
(From Ref. \protect\cite{harold-pier3}.)}
  \label{wtdir}
\end{figure}

Fig. \ref{wtdir} shows the probability distribution of the conductance [which is proportional to the total transmission coefficient, according to Landauer's formula, Eq. (\ref{landauer a})] for a
quantum chaotic billiard with one open channel in each lead ($N=1$), embedded in a magnetic field: the relevant universality class is thus the unitary one ($\beta =2$) (reported from Refs.
\cite{mello_kumar_oup,harold-pier3,wrm}).
The numerical conductance distribution was obtained collecting statistics by:

i) sampling in an energy window $\Delta E$ larger than the energy correlation length, but smaller than the interval over which the prompt response changes,
so that $\langle S \rangle$ is approximately constant across the window $\Delta E$.
Typically, 200 energies were used inside a window for which
$kW/\pi \in [1.6,1.8]$ (where $W$ is the width of the lead).
That such energy scales can actually be found in the physical system under study is interpreted as giving evidence of two rather widely separated time scales in the problem.

ii) sampling over 10 slightly different structures, obtained by changing the height or angle of the convex ``stopper", used mainly to increase the statistics.

Since we are for the most part averaging over energy, we rely on ergodicity to compare the numerical distributions with the ensemble averages of the theoretical maximum-entropy model. The optical $S$ matrix was extracted directly from the numerical data and used as $\langle S \rangle$ in Eq. (\ref{poisson}) for Poisson's kernel;
in this sense the theoretical curves shown in the figure represent
{\em parameter-free} predictions.

Several physical situations were considered in order to vary the amount of direct processes, quantified through $\langle S \rangle$.
The upper panels in Fig. \ref{wtdir} correspond to structures with leads attached outside the cavity, whereas the lower panels show the conductance distribution when the attached leads are extended into the cavity;
in the latter case, the presence of direct transmission is promoted.
In some instances, a tunnel barrier with transmission coefficient 1/2 has been included in the leads (indicated with dashed lines in the sketches of the cavities in the figure), in order to cause direct reflection.
The magnetic field was increased as much as to produce a cyclotron radius smaller than a typical size of the cavity, and about twice the width of the leads. The cyclotron orbits are drawn to scale on the left of Fig. \ref{wtdir} for low and high fields.
In all cases, the agreement between the numerical solutions of the
Schr\"{o}dinger equation and our maximum-entropy model is, generally
speaking, found to be good \cite{harold-pier3}.
We should remark that in Figs. 3(e) and 3(f) four subintervals (treated independently) of 50 energies each had to be used, as each subinterval showed a slightly different optical $S$.
This observation will be relevant to the discussion to be presented below on the validity and applicability of the information-theoretic model.

\begin{figure}
  \centerline{
    \includegraphics[width=.8\textwidth]
    {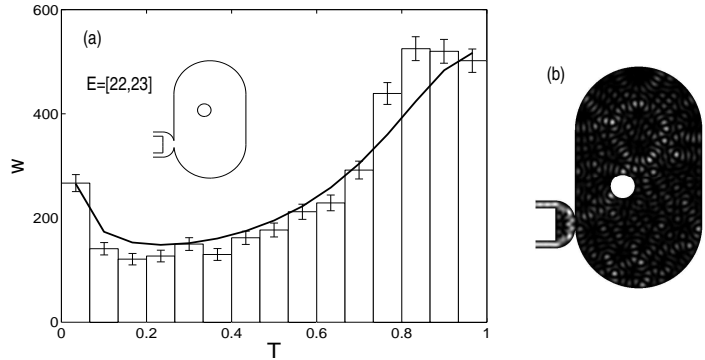}}
  \caption{(a) The conductance distribution obtained from the numerical integration of the Schr\"odinger equation for the structure shown in the inset and described in the text, compared with the prediction of Poisson's kernel, Eq. (\ref{poisson}), with
$\left\langle S \right\rangle$ extracted from the numerical data.
The system is time-reversal invariant, so that we are dealing with the universality class $\beta =1$.
The agreement is reasonable, although one observes systematic deviations, as evidenced by the statistical error bars.
(b) The square of the scattering wave function for a fixed energy inside the system considered in the analysis. We interpret the concentration of the wave function along the wall of the small cavity as a ``whispering gallery mode".
(From Ref. \protect\cite{bulgakov}.)}
  \label{wgm_vs_poisson 1}
\end{figure}

In order to elaborate further on the last remark, we investigated more closely a number of systems where the conditions for the approximate validity of stationarity and ergodicity are not fulfilled.
Fig. \ref{wgm_vs_poisson 1}(a) shows the statistical distribution of the conductance for a structure consisting of half a Bunimovich stadium coupled to a larger stadium.
The system is time-reversal invariant, so that we are dealing with the universality class $\beta =1$.
The smaller stadium supports a ``whispering gallery mode"
--providing the short path in this problem--
as illustrated in Fig. \ref{wgm_vs_poisson 1}(b), which shows the square of the absolute value of the scattering wave function for a fixed energy.
The larger stadium provides a ``sea" of fine structure resonances and should be responsible for the longer time scale in the problem.
The two waveguides support $N=1$ open channel each and the system is time-reversal invariant ($\beta=1$) (Ref. \cite{bulgakov}).
Statistics were collected by:

i) sampling in an energy window $\Delta E$.
It was found that even for $\Delta E$'s containing only $\approx 20$ points
--for which the $S(E)$'s were approximately uncorrelated--
the energy variation of $\langle S(E) \rangle$ inside $\Delta E$ was not negligible.

ii)
constructing an ensemble of 200 structures by moving the position of the
circular obstacle (shown in Fig. \ref{wgm_vs_poisson 1}) located inside the larger stadium.

Again, the value of the optical $S$ was extracted from the data and used as an input for Poisson's kernel, Eq. (\ref{poisson}).
We observe that the agreement between theory and the computer simulation is
reasonable.
However, judging from the statistical error bars (and the results for other intervals as well), the discrepancies shown in the figure seem to be systematic.
We interpret the situation described in i) above as giving evidence of two not very widely separated time scales in the problem.

\begin{figure}
  \centerline{
    \includegraphics[width=.8\textwidth]
    {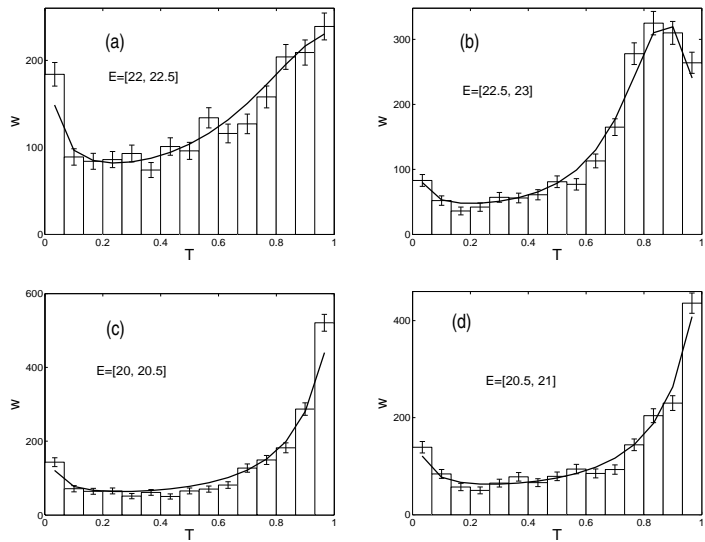}}
  \caption{The conductance distribution for the same structure as in
Fig. \ref{wgm_vs_poisson 1}, but using energy intervals twice as small for the construction of the histograms.
Panels (a) and (b) show the same data of Fig. \ref{wgm_vs_poisson 1}, but analyzed inside each of the two subintervals.
Panels (c) and (d) show the data for two other similar subintervals.
The agreement with theory has improved considerably.
(From Ref. \protect\cite{bulgakov}.)}
 \label{wgm_vs_poisson 2}
\end{figure}

With the idea of decreasing the energy variation of $\langle S(E) \rangle$ inside the energy interval $\Delta E$ used to construct the histogram,
$\Delta E$ was divided into two equal subintervals, giving the two conductance distributions shown in Figs. \ref{wgm_vs_poisson 2}(a) and (b).
We notice the improvement in the agreement between theory and numerics.
The results for two other similar subintervals are shown in
Fig. \ref{wgm_vs_poisson 2}(c) and (d), where a similar conclusion applies.
We remark that, in contrast to the results shown in Fig. \ref{wtdir}, now we are ``mainly" averaging over the ensemble of obstacles described in ii) above.

We wish to remark that, no matter how much $\Delta E$ needs to be decreased in order to reduce the energy variation of $\langle S(E) \rangle$ across it, the improvement between theory and experiment would not be surprising if the number of resonances inside $\Delta E$ were kept constant, so as to keep approximate stationarity.
This could be achieved, for instance, by increasing the size of the big cavity.
However, this is not what has been done: the size of the big cavity has been kept fixed, so that decreasing $\Delta E$ the number of resonances inside it decreases, and stationarity and ergodicity are ever less fulfilled. Yet, the agreement improves considerably: this we find surprising.

The results of a recent calculation 
\cite{victor-antonio}
are even more intriguing.
The conductance distribution was obtained numerically for the same structure as in the previous two figures, but now reducing the energy window $\Delta E$ literally to a point
--so that the distribution actually represents statistics collected across the ensemble at a fixed energy--
and the histogram was compared with the theoretical model.
The results are shown in Fig. \ref{N=1,DeltaE=0} for two different energies: they show a very good agreement with theory.
\begin{figure}
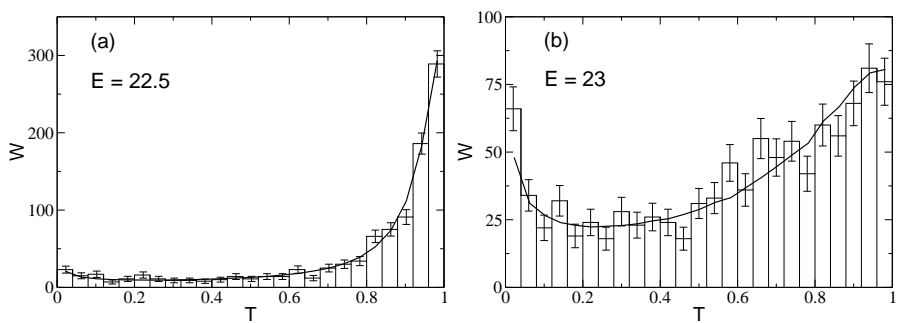

  \centerline{
    \includegraphics[width=.5\textwidth]{mello_fig6a.eps}
    \includegraphics[width=.5\textwidth]{mello_fig6b.eps}}
  \caption{The conductance distribution for the same structure as in
Figs. \ref{wgm_vs_poisson 1} and \ref{wgm_vs_poisson 2}, but using an energy window $\Delta E=0$ for the construction of the numerical histogram, which then represents statistics collected across the ensemble of obstacles at a fixed energy. 
Panels (a) and (b) correspond to two different energies.
The agreement with theory is, in each case, very good.
}
  \label{N=1,DeltaE=0}
\end{figure}

\begin{figure}
  \centerline{
    \includegraphics[width=.5\textwidth]{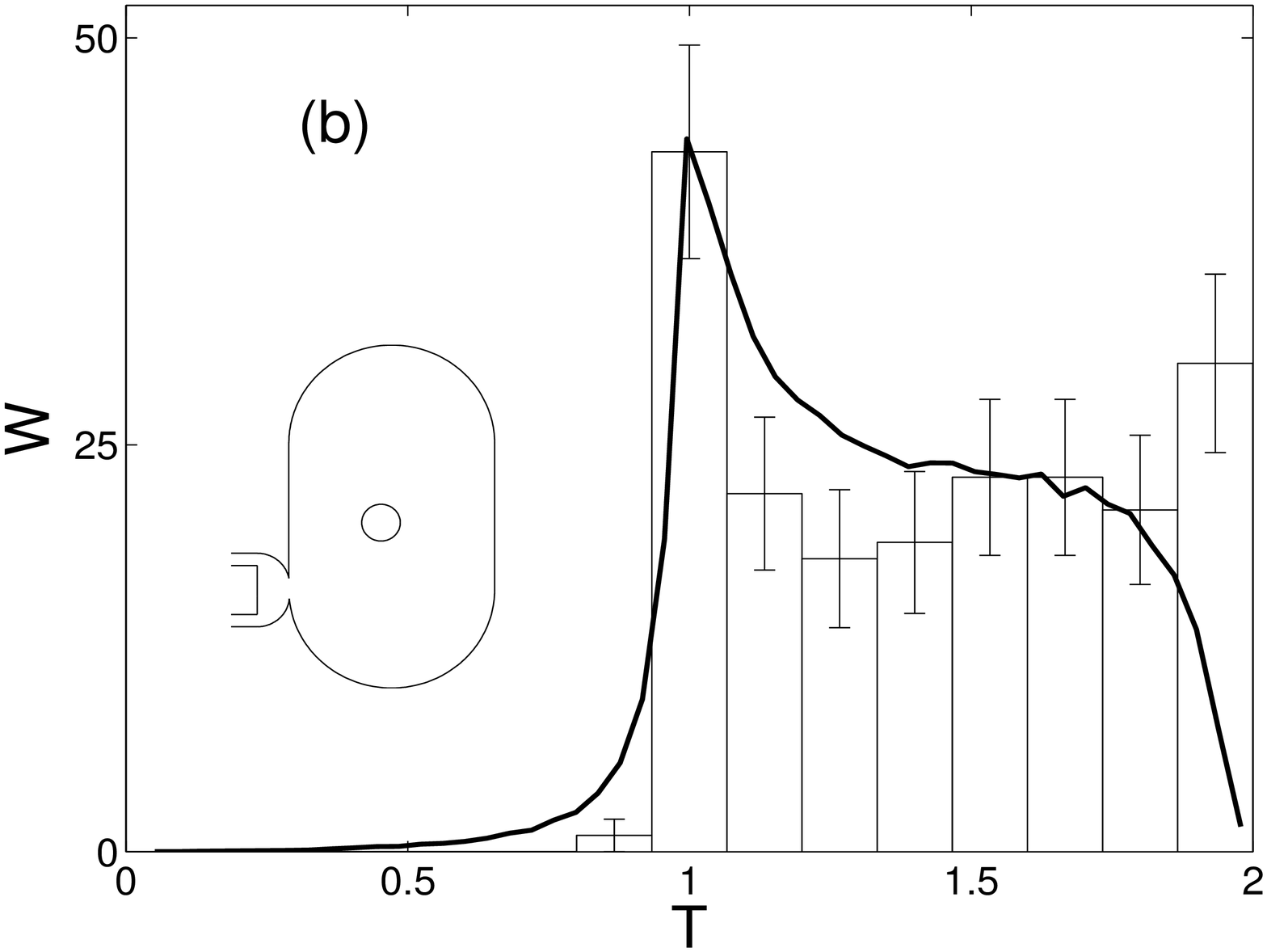}}
  \caption{The conductance distribution for the structure shown in the inset, at an energy such that $N=2$ channels are open.
The system is time-reversal invariant, so that we are dealing with the universality class $\beta =1$.
The energy interval $\Delta E$ was literally reduced to a point (after Ref. \protect\cite{bulgakov}).
The agreement between theory and the simulation is reasonable.}
  \label{wgm_vs_poisson N=2}
\end{figure}

In Fig. \ref{wgm_vs_poisson N=2} we present the conductance distribution for the same two-stadium structure as in the previous figures, again for the universality class $\beta=1$, but now at an energy such that the waveguides support $N=2$ open channels. Just as in Fig. \ref{N=1,DeltaE=0}, the numerical analysis was done collecting data across the ensemble for a fixed energy. The agreement, although not perfect, is reasonable.

\subsection{Discussion}
\label{discussion}

The theoretical model that we have used to compare with computer simulations, i.e., Poisson's kernel of Eq. (\ref{poisson}), is based on the properties of analyticity, stationarity and ergodicity, plus a maximum-entropy ansatz.
The analytical properties of the $S$ matrix are of general validity.
On the other hand, ergodicity is defined for stationary random processes, and
stationarity, in turn, is an extreme idealization that regards $S(E)$ as a stationary random (matrix) function of energy: stationarity implies, for instance, that the optical
$\langle S(E) \rangle$ is constant with energy and the characteristic time associated with direct processes is literally zero.
Needless to say, in realistic dynamical problems like the ones considered in the present section, stationarity is only approximately fulfilled, so that one has to work with energy intervals $\Delta E$ across which the ``local" optical
$\langle S(E) \rangle$ is approximately constant, while, at the same time, such intervals should contain many fine-structure resonances.
Above we saw examples in which this compromise is approximately fulfilled, and cases in which it is not.
Yet, we have given evidence that reducing the size of $\Delta E$, even at the expense of decreasing the number of resonances inside it, improves significantly the agreement between theory and numerical experiments.
We went to the extreme of reducing $\Delta E$ literally to a point, which amounted to leaving the energy fixed and collecting data over the ensemble constructed by changing the position of the obstacle: in the example shown in Fig. \ref{N=1,DeltaE=0} we found an excellent agreement of the resulting distribution with Poisson's kernel.

An essential ingredient to construct Poisson's kernel of Eq. (\ref{poisson}) is the reproducing property of Eq. (\ref{repr_property}). This is a property of the ensemble, defined for a {\em fixed energy}: at present we only know how to justify it through the properties of stationarity and ergodicity.
However, the results we have shown
(Fig. \ref{wgm_vs_poisson 2}, which is ``almost" an ensemble average, and Fig. \ref{N=1,DeltaE=0} which is literally an ensemble average), seem to suggest that Poisson's kernel is valid beyond the situation where it was originally derived: i.e., it is as if the reproducing property of Eq. (\ref{repr_property}) were valid even in the absence of stationarity and ergodicity.
We do not have an explanation of this fact at the present moment.

\section{Conclusions}
\label{conclusions}

In this contribution we reviewed a model that was originally developed
in the realm of nuclear physics, and applied it to the description of statistical Quantum Mechanical scattering of one-particle systems inside cavities whose underlying classical dynamics is chaotic.
The same model is also applicable to the study of scattering of classical waves by cavities of a similar shape.

In the scheme that we have described, the $S(E)$ matrix is idealized  as a stationary random (matrix) function of $E$, satisfying the condition of ergodicity. It is assumed that we have two distinct time scales in the problem, arising from a prompt and an equilibrated component: the former one is associated with direct processes, like short paths, and is described through the optical matrix
$\langle S \rangle$.

The statistical distribution of $S$ is proposed as one of maximum entropy, once the reproducing property described in the text is fulfilled and the system-dependent optical $\langle S \rangle$ is specified.
In other words, system-specific details other than the optical matrix $\langle S \rangle$ are assumed to be irrelevant.

The predictions of the model for the conductance distribution in the presence of direct processes should be experimentally observable in microstructures where phase breaking is small enough, the sampling being performed by varying the energy or shape of the structure with an external gate voltage.
They have also been observed in microwave scattering experiments.

Comparison of the theoretical predictions with computer simulations indicate that the present formulation has a remarkable predictive power: indeed, it describes statistical properties of the quantum conductance of
cavities, like its average, its fluctuations, and its full
distribution in a variety of cases, both in the absence and in the presence of a prompt response. Here we have concentrated on the conductance distribution, mainly in the presence of prompt processes.

In Ref. \cite{harold-pier3} we found that, with a few exceptions, it was possible to find energy intervals $\Delta E$ containing many fine-structure resonances in which approximate stationarity could be defined; the agreement was generally found to be good.
In contrast, in Ref. \cite{bulgakov} we encountered cases in which this was not possible and we undertook a closer analysis of the problem.
We found that reducing $\Delta E$, even at the expense of decreasing the number of fine-structure resonances inside it, improved the agreement considerably. In a number of cases, $\Delta E$ was reduced literally to a point and the data were collected over an ensemble constructed by changing the position of an obstacle inside the cavity: in other words, Poisson's kernel was found to give a good description of the statistics of the data taken across the ensemble for a fixed energy.
As a result, Poisson's kernel seems to be valid beyond the situation where it was originally derived, where the properties of stationarity and ergodicity played an important role.
It is as though the reproducing property of Eq. (\ref{repr_property}),
which is a property of the ensemble defined for a fixed energy, were valid even in the absence of stationarity and ergodicity, which were originally used to derive it.
At the present moment we are unable to give an explanation of this fact.


\end{document}